\documentclass[prl,showpacs,twocolumn]{revtex4}
\usepackage{graphicx}
\usepackage{amsmath}
\usepackage{amssymb}
\usepackage{color}
\usepackage[normalem]{ulem}
\usepackage{epsfig}

\begin{document}
\title{Fronts and fluctuations at a critical surface}
\author{Haim Weissmann, Nadav M. Shnerb and David A. Kessler }
\affiliation{Department of Physics, Bar-Ilan University, Ramat-Gan
IL52900, Israel} \pacs{87.10.Mn,87.23.Cc,64.60.Ht,05.40.Ca}
\begin{abstract}
The properties of a front between two different phases in the presence of a smoothly inhomogeneous external field
that takes its critical value at the crossing point is analyzed. Two generic scenarios are studied. In the first, the system admits a bistable solution and the external field governs the rate in which one phase invades the other. The second mechanism corresponds to a second order transition that, in the case of reactive systems, takes the form of  a transcritical bifurcation at the crossing point. We solve for the front shape and its response to external white noise, showing that static properties and also some of the dynamics features cannot distinguish between the two scenarios. The only reliable indicator turns out to be the fluctuation statistics. These take a Gaussian form in the bifurcation case and a double-peak shape in a bistable system. The results of a recent analysis of the morphogenesis process in \emph{Drosophila} embryos are reanalyzed and we show, in contrast with the interpretation suggested by Krotov, et al. \cite{krotov2014morphogenesis}, that the plausible underlying dynamics is bistable and not bifurcational.
\end{abstract}
\maketitle

The theory of first and second order phase transitions is a well-established part of
statistical physics, and its generalization to out-of-equilibrium problems, like glassy behavior and
percolation-like transitions, has also received a lot of attention during the last decades. However, most of these works  focused on the case where the control variable (like temperature) is homogenous in space. Only recently has the equilibrium properties of thermodynamic systems with a spatially varying temperature that takes its critical value only in a localized region begun to be studied~\cite{bonati2014universal}.

   Coincidentally, an out-of-equilibrium process with the same spatial characteristics  was suggested as the underlying mechanism behind one of the most fundamental and universal aspects of developmental biology, embryonic morphogenesis. Measuring the expression levels of gap genes along the anterior-posterior axis of \emph{Drosophila} embryos, Krotov, et al. \cite{krotov2014morphogenesis} revealed a strong negative correlation between spatial expression levels of gene pairs, with an increase in the expression level of gene A along the spatial axis  associated with a decrease in expression of gene B. Krotov, et al. interpret their findings as evidence for competition between A and B; i.e., they assume that the transcription activity of A is reduced  when gene B is active and vice versa. In the simplest two-gene model, the primary maternal morphogens enhance the production of A to the left of a crossing point and the production of B to its right. At the crossing point, where there is no preference to any of the genes, the system is at ``criticality", showing larger fluctuations with a fast manifold associated with conservation of the global activity and a slow manifold that corresponds to the differences between concentrations of A and B. To model this process, Krotov, et al. implemented a simple reaction-diffusion model with a spatial gradient of the control parameter (maternal morphogens) and white noise, arguing that the analytical and numerical results obtained from this model  fit the empirically observed data.

However, we believe there is a loophole in the logic of Krotov, et al. Their model, which involved two coupled logistic populations with a  transcritical bifurcation at the crossing point is indeed one generic possibility, but there is an alternative generic scenario, wherein the system is bistable and the external field reaches the stall point at the transition. In this letter we analyze these two scenarios and solve for the front width and the fluctuation spectrum at the crossing point. Doing that, one realizes that the second scenario, the bistable front model, better captures the features of the morphogenetic process considered in \cite{krotov2014morphogenesis}.

Qualitatively speaking, the bistable scenario is the analog of a simple first order transition system, although (as we shall see below) one should make a distinction between its equilibrium and out-of-equilibrium properties. Imagine a water-ice mixture in three dimensions, say, where the temperature depends on $x$ and $T(x=0)=T_m$, where $T_m$ is the melting temperature. In the right half-space water invades ice and in the left ice grains grow in water, so the ice-water front will be trapped around $x=0$. Due to thermal or other fluctuations the front will move back and forth in a region determined by $x_t$, leading to a characteristic fluctuation spectrum at $x=0$; below we will distinguish between the instantaneous shape of the front and its average over time and will analyze the distribution of fluctuations.

The bifurcation model, which is the one considered in \cite{krotov2014morphogenesis}, is slightly more complicated. This model is a nonequilibrium continuous (``second order") transition, with a transcritical bifurcation at $x=0$. To imitate the gene expression case one needs two competing fields. A generic set of PDEs that yields the required behavior is
\begin{eqnarray} \label{eq1}
\frac{\partial a(x,t)}{\partial t} = D \nabla^2 a + a - \frac{a^2+[1+C(x)]ab}{K} + \zeta_a(x,t) \nonumber \\
 \frac{\partial b(x,t)}{\partial t} = D \nabla^2 b + b - \frac{b^2+[1-C(x)]ab}{K} +\zeta_b(x,t).
 \end{eqnarray}
 Where $a$ and $b$ are the expression levels of the corresponding genes A and B, $C(x)$ is the background field (maternal morphogenes) that switches sign at zero and $\zeta$ is, say,  white noise.

  In the absence of noise, $\zeta=0$, in the regime $C(x)>0$ the fixed point (FP) of this system  is $a=0, \ b=1$ while if $C(x)<0$ the FP is $a=1, \ b=0$. The diffusion term, however, couples the left and the right regions, and the expression level must be a smooth function of $x$ that approaches the FPs at large $|x|$ and takes the symmetric value $a=b=1/2$  at the crossing point.

  An appropriate and quite generic choice of the external field  profile is $C(x) = \tanh(x/x_t)$, where $x_t$ sets the scale of the external field gradient. Plugging $b = 1-a$ into Eqs. (\ref{eq1})  one finds in the deterministic limit (from here on we normalize $K$ to unity), ${\dot a} = D \nabla^2 a - C(x) a (1-a)$; expanding $C$ at $x\ll x_t$ the equation for the stable front is then
  \begin{equation} \label{eq2}
 a''(y) - y a(y) [1-a(y)] = 0,
  \end{equation}
  where $y \equiv x (Dx_t)^{-1/3}$ and spatial derivatives are taken with respect to $y$.

  Eq. (\ref{eq2}) describes  a deterministic, spatial voter model with selection \cite{Korolev2010genetic}, where the value of $y$ determines the preference towards one of the "alleles" (opinions). The front width is proportional to $[Dx_t]^{1/3}$; its shape and the fluctuation spectrum at the front will be determined below.

\begin{figure}
\centering{
\includegraphics[width=8cm]{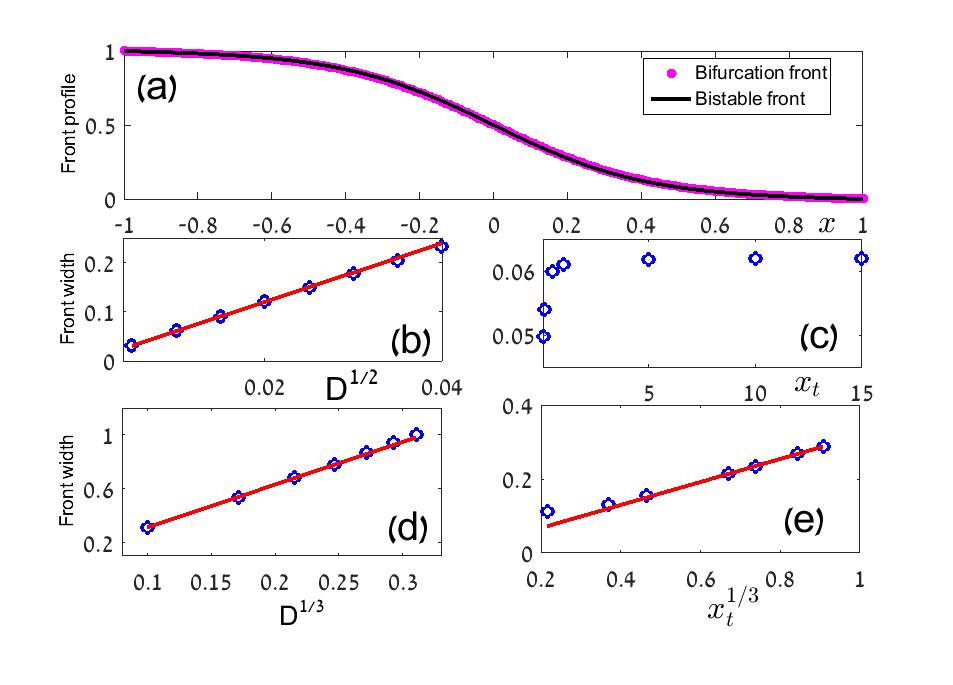}}
\caption{The static properties of the deterministic fronts:  panel (a) shows both static front profiles, $a_0(x)$ (pink, wider) and $\phi_0(x)$ (black line), for $D=0.4$ and $x_t = 0.5$. Clearly there is no essential difference between the two. In panel (b)  the width of the bifurcation front was plotted against $D^{1/2}$ (for fixed $x_t = 1$, results are open circles, red line is a linear fit) and in panel (d) the same quantity is plotted against
$x_t$ for fixed $D=10^{-4}$; as predicted, the bistable front width is linear in the square root of $D$ and in independent of $x_t$ for when $x_t$ is larger than the natural width of the front. Panels (d) and(e) show the same relations for a bifurcation front ($D$ varies for $x_t=1$, $x_t$ varies for $D=10^{-3}$), demonstrating the $(Dx_t)^{1/3}$ scaling.    } \label{fig1}

\end{figure}

  It appears to be instructive to contrast the bifurcation model (\ref{eq2}) with the other generic scenario, a bistable system. Let us consider the simplest model of a bistable system with a crossing point, described by  a spatially inhomogeneous  Ginzburg-Landau (GL) Equation:
  \begin{equation} \label{eq3}
   {\dot \phi}(x,t) = D \nabla^2 \phi(x,t) + \phi(x,t) [1-\phi(x,t)][\phi(x,t)-C(x)].
  \end{equation}
  When $C$ is a  constant, $\phi$ admits three spatially homogenous FPs, two stable FPs at $\phi = 0$ and $\phi=1$ and an unstable FP at C, the $\phi =1$ invades the zero phase if $C <1/2$ and zero invades if $C>1/2$. $C=1/2$ is the melting point, or the stall point of the GL front; at the melting point, when the system evolves from inhomogeneous initial conditions like $\phi = 0$ for $x<0$ and $\phi=1$ for $x>0$, it relaxes to the stable front solution, $\phi_0 = [\tanh(x/\sqrt{8D})+1]/2$. Accordingly, even when $C$ is $x$ dependent, as in the case  $C(x) = \tanh(x/x_t)$ considered above, as long as the intrinsic width of the front $\sqrt{8D}$ is much smaller then $x_t$, the shape and the width of the front will be essentially independent of the external field (see Figure \ref{fig1}). Although both scenarios support a stable front, the dependence of its width on the external parameters is different: in a bifurcation system this width scales like $D^{1/3}$  and depends on the width of the crossing region $x_t$, while in a bistable system the width scales like $D^{1/2}$  and is independent of $x_t$ when  $x_t$ is large.

  However, when a front is observed, as in the experiment discussed by \cite{krotov2014morphogenesis}, and one would like to determine the underlying mechanism, the utility of diagnostic tools based on static properties of the front is quite limited. In experiments it is quite difficult to change $D$ or $x_t$ - to do that, the dynamics on the molecular level should be manipulated - so one cannot measure the dependency of the front width on $D$. Worse than that, it turns out that the front shape is almost identical in both cases. In the bifurcation model $\phi_0 \equiv 1/2 + x/\sqrt{16D}$ close to $x=0$, meaning that the front satisfies, to  first order in $x$ and $\phi$, $ D \nabla^2 \phi(x) + (x/\sqrt{16D}) \phi(x) [1-\phi(x)]=0$, i.e., it is equivalent, up to a constant, to the bifurcation front solution (\ref{eq2}), so the differences between $\phi_0(x)$ and the solution of Eq. (\ref{eq2}) (denoted hereon as $a_0(x)$) are extremely small, as demonstrated in Figure \ref{fig1}.  Without measuring the diffusion constant of the underlying morphogenesis molecular agents, or monitoring the front profile to a very high degree of accuracy, one cannot use static properties to distinguish between the two possible scenarios.

Not only the static properties of the front are practically useless as an indicator of the underlying mechanism, the same is true for some dynamical aspects of the dynamics.  In \cite{krotov2014morphogenesis}, for example, the location of the crossing point was identified (assuming an underlying bifurcation scenario) by a peak in the anticorrelations between the densities of $a$ and $b$, and the existence of a slow and fast manifold was demonstrated by a scatter plot of the fluctuations, showing that the sum $a+b$ is kept almost fixed through time but the differences $a-b$ fluctuate strongly. Indeed, the same features are also a characteristic of a bistable scenario. To show that, we have developed a two-species model that supports bistability (in the one species case (\ref{eq3}) the features demonstrated in \cite{krotov2014morphogenesis} are embarrassingly trivial, since the field $\phi$ should be interpreted such that $a \equiv \phi$ and $b \equiv 1-\phi$, so the sum is fixed and the anticorrelations are guaranteed in advance).

\begin{figure}
{\centering
\includegraphics[width=8cm]{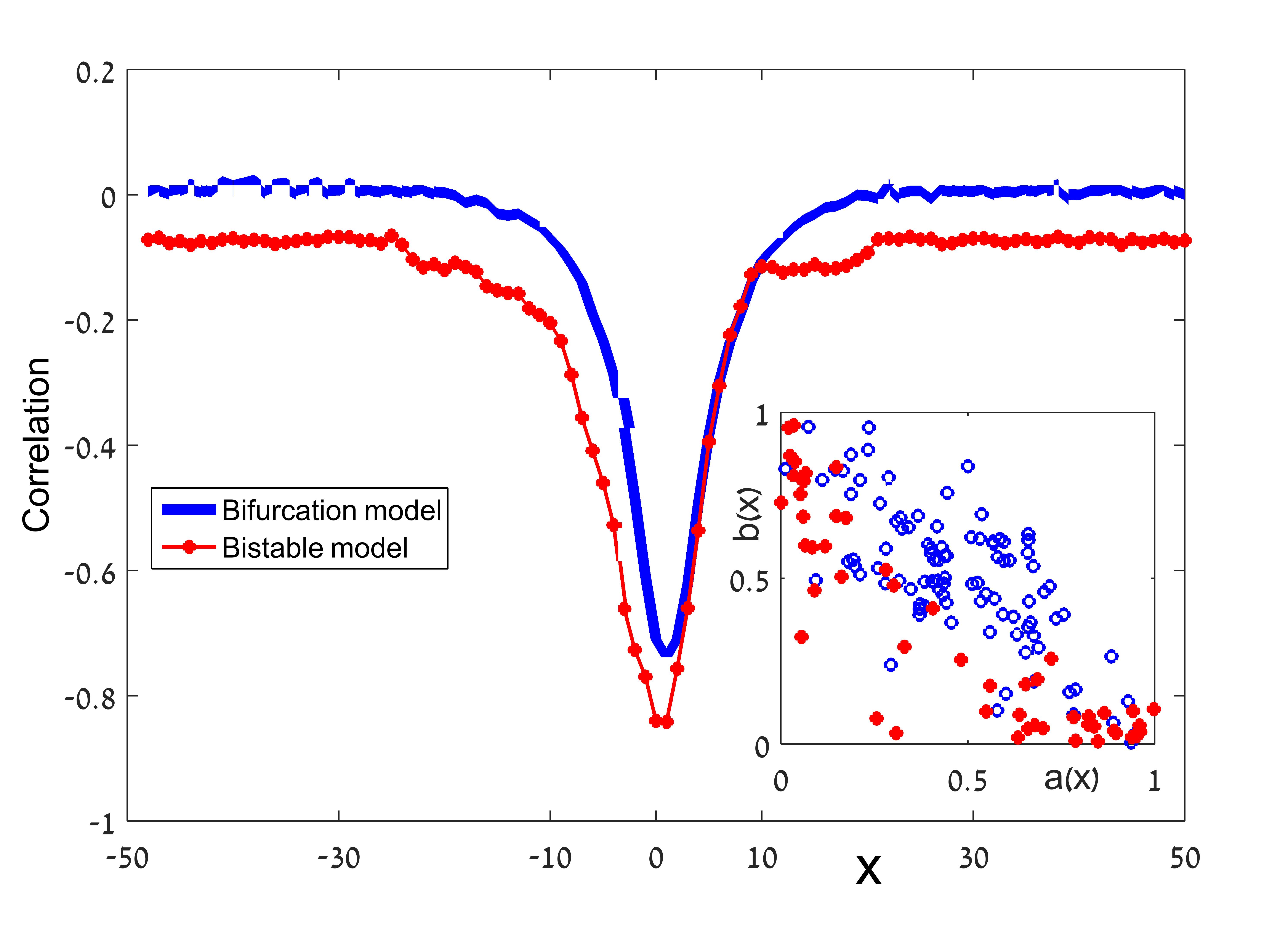}
}
\caption{The correlation between two timeseries, $a(x,t)$ and $b(x,t)$, for a fixed distance  $x$ from the crossing point, is plotted against $x$ for both bifurcation model (full blue line) and the bistable model (red line with full circles). In both models the anticorrelations between  fluctuations of $a$ and $b$ are peaked at $x=0$ (main panel). In the inset, a scatter plot of the instantaneous fields at different times is presented in the $a-b$ plane (without fluctuations there should be one point at $1/2,1/2$). Points from the bistable model are represented by filled red circles, bifurcation model points are empty and blue. Fast ($a+b$) and slow ($a-b$) manifolds are clearly seen. Results were obtained with $x_t =10, \ D=1/2$ and noise amplitude $0.04$.      } \label{fig2}
\end{figure}

To construct a  simple two species bistable model we define $S(x,t) = a(x,t) +b(x,t)$ and $Q(x,t) = a(x,t) - b(x,t)$, and the local dynamics satisfies
\begin{equation} \label{eq4}
{\dot S} = S(\alpha -S) \qquad {\dot Q} = (Q-C(x)) (S^2 - Q^2)
\end{equation}
so the stable FPs correspond to $S=\alpha$ and $Q=\pm \alpha$ and for constant $C$ there is an unstable FP at $Q=C$. Both reactants $a$ and $b$ diffuse with a diffusion constant $D$. As shown before, the bistable front has an intrinsic width and its shape is independent of the external field parameter $x_t$ as long as $x_t \gg \sqrt{8D}$. Accordingly we simulate this system with antiperiodic boundary conditions and without an external field. The correlation function of $a$ and $b$, together with a scatter plot of the fluctuations at the crossing point, are depicted in Figure \ref{fig2}, showing that both models have very similar qualitative behavior.

However, the sharp-eyed observer may notice a subtle qualitative difference between the scatter plots of fluctuation amplitudes. In the bifurcation model simulations the points appear to have higher density in the middle (around $[0.5,0.5]$, which is the steady state value of the front at the crossing point), while the simulation of the bifurcation model yields higher concentration of fluctuation points close to the two extremes $a=0, \ b=1$ and $a=1, \ b=0$. This is not an accident, and  provides  a crucial hint: the two  mechanisms, bifurcation and bistability, admit qualitatively different fluctuation statistics. In the bistable scenario, due to the absence of the external gradient, the noise causes the front to move back and forth freely around the crossing point, so at $x=0$ the system is almost always either it the $\phi=1$ state or the $\phi=0$ state, leading to a fluctuation spectrum with two peaks at zero and one and a dip at $1/2$. The bifurcation mechanism, on the other hand, yields only a single peak around the steady state value $a_0(x=0)=b(x=0)=1/2$.

To quantify this, we consider first the fluctuations around the steady state front of the bifurcation model,  $a_0(x) + \delta(x,t)$ in the presence of an external white noise. Linearizing Eq. (\ref{eq2}) to the first order in $\delta$, and taking into account the front shape close to the crossing point,  $a_0(x) \sim 1/2 + c_1 x/(Dx_t)^{1/3}$, where $c_1$ is an ${\cal O}(1)$ constant, one obtains a dynamical equation for the fluctuations of the bifurcation model:
\begin{equation} \label{eq5}
{\dot \delta}(x,t) = D \nabla^2 \delta(x,t) - \kappa x^2 \delta(x,t) + \zeta(x,t)
\end{equation}
where $\kappa = c_1 (D^{1/3}x_t^{4/3})^{-1}$ and $\zeta$ is a white noise, $\overline{\zeta(x,t)}=0$ and $\overline {\zeta(x,t)\zeta(x',t')}=\Delta \delta(x-x')\delta(t-t')$ where an overbar represent an average over realizations. Expanding $\delta$ in terms of normalized quantum harmonic oscillator wavefunctions, $\delta(x,t) = \sum \beta_m(t) \psi_m(x)$, and using their orthonormality properties one obtains,
\begin{equation} \label{eq6}
{\dot \beta_m(t)} = -\Gamma_m \beta_m(t) + \eta(t),
\end{equation}
with $\Gamma_m = 2\sqrt{D \kappa}(m+1/2)$ and $\eta(t)$ is, again, white noise. Every coefficient $\beta_m$ is subject to an Ornstein–Uhlenbeck process and its probability distribution function is given by a Gaussian with zero mean and variance $\Delta/\Gamma_m$. An immediate result is that $\delta$ itself is a zero mean Gaussian, i.e., that the fluctuation density histogram is a Gaussian centered at $a_0(x=0)=1/2$. Indeed one can do even better and calculate the variance of this Gaussian,
\begin{eqnarray} \label{eq6}
&Var(\delta) = \sum \limits_{m \ {\rm even}} \psi^2_m(0) Var(\Gamma_m)   \\ \nonumber &=  \frac{\Delta}{2\sqrt{\pi}c_1^{3/8}} \sqrt{\frac{x_t}{D}} \sum \limits_{m \ {\rm even}} \frac{\left((m-1)!!\right)^2}{m!(m+1/2)} = \frac{\Delta \Gamma(5/4)}{\Gamma(3/4) c_1^{3/8}} \sqrt{\frac{x_t}{D}}.
\end{eqnarray}

\begin{figure}
\vspace{-2.5cm}
\includegraphics[width=7cm]{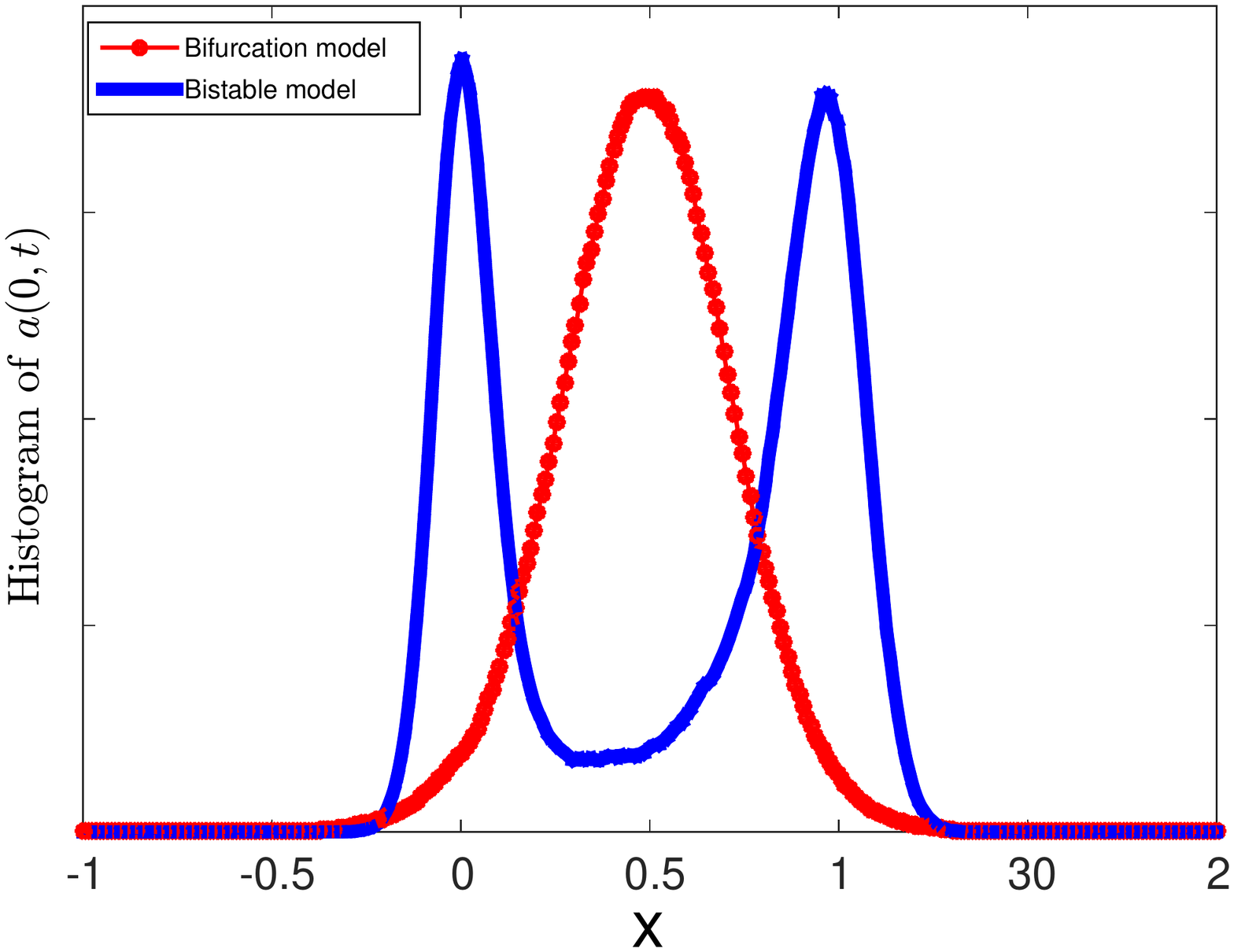}
\vspace{-2.5cm}
\caption{Fluctuations statistics: a histogram (unnormalized) of $a(t)$ values at the crossing point for the bifurcation (red line with filled circles) and bistable (blue) models. In both cases noise leads to deviations from the steady state value $a = 0.5$, however in the bifurcation case these deviations are distributed normally around the average while the bistable system distribution is bimodal. Simulation parameters are identical to those specified in the caption of Figure \ref{fig2}.} \label{fig3}
\end{figure}

In a bistable system the situation is completely different. As explained above,  as long as $x_t$ is significantly larger than the internal width of the front, one can replace the external field (with exponentially small corrections in a finite system) by antiparallel boundary conditions at $\pm \infty$, and the   the fluctuations admit a zero (Goldstone) mode since the location of the front is translationaly invariant. Accordingly, one finds the crossing point either in the $a$ phase or in the $b$ phase, with fluctuations due to the effect of noise on any of these phases. As a result the histogram of fluctuations amplitude, instead of being a Gaussian around $1/2$, has two peaks that correspond to the two attractive fixed points of the homogenous problem. These features are demonstrated in Figure \ref{fig3}, where the strong qualitative difference, allowing for easy discrimination between the two scenarios, is manifest. On the other hand, when $x_t$ is much smaller than the natural width of the front, even the bistable system loses its translational invariance property, the front is trapped by the external field and cannot change significantly its spatial location, and the resulting fluctuation spectrum is peaked at $1/2$. In such a case we cannot offer a simple method to distinguish between the two alternatives mechanisms.

Turning back to the work of  Krotov, et al. \cite{krotov2014morphogenesis}, the results from the experiment they analyzed clearly show a crossing regime with anticorrelations between $a$ and $b$ with fast and slow manifold, however, as we explained here, this cannot reveal the nature of the dynamics governs the system. The only simple qualitative indicator is the histogram of the amplitude of the fluctuations, and their results (Fig 3C of \cite{krotov2014morphogenesis}) clearly show a double peak structure, meaning that the underlying dynamics is evidently bistable, equivalent to a first order transition with an external field (primary maternal morphogenes) that changes the ``temperature" such that the melting temperature marks the crossing point. This appears to rule out the bifurcational interpretation suggested in \cite{krotov2014morphogenesis}.

Finally, in the context of the bistable model we would like to stress the difference between two possible definitions of a front separating two phases. The analysis followed Eq. (\ref{eq3}) regards the \emph{instantaneous} shape of a front, i.e., the typical shape of a snapshot of the crossover region. In contrary one may define the \emph{time averaged}, or the "equilibrium" front, wherein that the $a$ density, say, is averaged at $x$ over a long time span and the resulting front is  the profile of $\langle a(x)\rangle$, where $\langle ...\rangle$ represents the time, or equivalently the thermodynamic, average.

The width of an equilibrium front under smoothly varying external field was analyzed by \cite{bonati2014universal} in the context of a 2d $q$-state Potts model, these authors found that for $q\ge 4$ when the system has a first-order transition, the width of the front $\langle a(x)\rangle$  scales as $x_t^{1/3}$. To understand and generalize their result, let us consider an equilibrium system at the transition point. Starting from a homogenous $A$ state, $B$ phase droplets with the same bulk energy are nucleated and shrink only due to the surface tension. As a result, the larger  the droplet, the more stable it is; monitoring the phase at a certain point $x$ for long time one finds $\langle a(x)\rangle = 1/2$, independent of the location of the measurement. Considering a randomly  moving  front, like the one described above, one arrives at the same conclusion.

What limits the size of these droplets, hence determining the width of the equilibrium front, is the external field gradient. If phase $A$ invades the region $x>0$ (where phase $B$ is prefered) by a compact semisphere of radius $A$, the bulk energy cost $U$ of such a droplet is,
\begin{equation} \label{eq8}
U \propto \int_0^R \left( R^2-x^2 \right)^{\frac{d-1}{2}} \frac{x}{x_t} dx \sim \frac{R^{d+1}}{x_t},
\end{equation}
meaning that the width of the equilibrium front scales like $x_t^{1/(d+1)}$ (the scaling $x_t^{1/3}$ when $d=2$, was found in \cite{bonati2014universal}). Accordingly, the width of the equilibrium front does depend on $x_t$, as opposed to the instantaneous front. In any case, the hallmark of a bistable system is the double peak of the fluctuation spectrum, not the shape of the front.

The problem considered here, a front pinned by smooth spatial gradient of an external field, appears to be quite generic. Beyond the experimental results that were considered in \cite{krotov2014morphogenesis}, it may be relevant to the effects of environmental gradient on the genetic heterozygosity  of a population (see, e.g., \cite{nanninga2014environmental}) and on the species richness, gene transfer and speciation  in ecological communities along such a gradient (known as an ecotone or ecoline) \cite{smith1997role,kark2013ecotones}. In particular, the distinction between a stable, bifurcational front and the wandering front characterizing a bistable scenario may be very relevant to the rate of gene flow and to the chance of ecotonal species to survive. Further studies of these phenomena, and in particular an appropriate classification of front dynamics using fluctuation statistics, may shed a new light on many fundamental processes both in physics and in the life sciences.

{\bf Acknowledgments} We thank Herbert Levine for helpful discussions. We acknowledges the support of the Israel
Science Foundation (NMS BIKURA grant no. $1026/11$, DAK no. $376/12$.

\bibliography{references_haim}

\begin{thebibliography}{6}
\expandafter\ifx\csname natexlab\endcsname\relax\def\natexlab#1{#1}\fi
\expandafter\ifx\csname bibnamefont\endcsname\relax
  \def\bibnamefont#1{#1}\fi
\expandafter\ifx\csname bibfnamefont\endcsname\relax
  \def\bibfnamefont#1{#1}\fi
\expandafter\ifx\csname citenamefont\endcsname\relax
  \def\citenamefont#1{#1}\fi
\expandafter\ifx\csname url\endcsname\relax
  \def\url#1{\texttt{#1}}\fi
\expandafter\ifx\csname urlprefix\endcsname\relax\def\urlprefix{URL }\fi
\providecommand{\bibinfo}[2]{#2}
\providecommand{\eprint}[2][]{\url{#2}}

\bibitem[{\citenamefont{Krotov et~al.}(2014)\citenamefont{Krotov, Dubuis,
  Gregor, and Bialek}}]{krotov2014morphogenesis}
\bibinfo{author}{\bibfnamefont{D.}~\bibnamefont{Krotov}},
  \bibinfo{author}{\bibfnamefont{J.~O.} \bibnamefont{Dubuis}},
  \bibinfo{author}{\bibfnamefont{T.}~\bibnamefont{Gregor}}, \bibnamefont{and}
  \bibinfo{author}{\bibfnamefont{W.}~\bibnamefont{Bialek}},
  \bibinfo{journal}{Proceedings of the National Academy of Sciences}
  \textbf{\bibinfo{volume}{111}}, \bibinfo{pages}{3683} (\bibinfo{year}{2014}).

\bibitem[{\citenamefont{Bonati et~al.}(2014)\citenamefont{Bonati, D'Elia, and
  Vicari}}]{bonati2014universal}
\bibinfo{author}{\bibfnamefont{C.}~\bibnamefont{Bonati}},
  \bibinfo{author}{\bibfnamefont{M.}~\bibnamefont{D'Elia}}, \bibnamefont{and}
  \bibinfo{author}{\bibfnamefont{E.}~\bibnamefont{Vicari}},
  \bibinfo{journal}{Physical Review E} \textbf{\bibinfo{volume}{89}},
  \bibinfo{pages}{062132} (\bibinfo{year}{2014}).

\bibitem[{\citenamefont{Korolev et~al.}(2010)\citenamefont{Korolev, Avlund,
  Hallatschek, and Nelson}}]{Korolev2010genetic}
\bibinfo{author}{\bibfnamefont{K.}~\bibnamefont{Korolev}},
  \bibinfo{author}{\bibfnamefont{M.}~\bibnamefont{Avlund}},
  \bibinfo{author}{\bibfnamefont{O.}~\bibnamefont{Hallatschek}},
  \bibnamefont{and} \bibinfo{author}{\bibfnamefont{D.~R.}
  \bibnamefont{Nelson}}, \bibinfo{journal}{Reviews of modern physics}
  \textbf{\bibinfo{volume}{82}}, \bibinfo{pages}{1691} (\bibinfo{year}{2010}).

\bibitem[{\citenamefont{Nanninga et~al.}(2014)\citenamefont{Nanninga,
  Saenz-Agudelo, Manica, and Berumen}}]{nanninga2014environmental}
\bibinfo{author}{\bibfnamefont{G.~B.} \bibnamefont{Nanninga}},
  \bibinfo{author}{\bibfnamefont{P.}~\bibnamefont{Saenz-Agudelo}},
  \bibinfo{author}{\bibfnamefont{A.}~\bibnamefont{Manica}}, \bibnamefont{and}
  \bibinfo{author}{\bibfnamefont{M.~L.} \bibnamefont{Berumen}},
  \bibinfo{journal}{Molecular ecology} \textbf{\bibinfo{volume}{23}},
  \bibinfo{pages}{591} (\bibinfo{year}{2014}).

\bibitem[{\citenamefont{Smith et~al.}(1997)\citenamefont{Smith, Wayne, Girman,
  and Bruford}}]{smith1997role}
\bibinfo{author}{\bibfnamefont{T.~B.} \bibnamefont{Smith}},
  \bibinfo{author}{\bibfnamefont{R.~K.} \bibnamefont{Wayne}},
  \bibinfo{author}{\bibfnamefont{D.~J.} \bibnamefont{Girman}},
  \bibnamefont{and} \bibinfo{author}{\bibfnamefont{M.~W.}
  \bibnamefont{Bruford}}, \bibinfo{journal}{Science}
  \textbf{\bibinfo{volume}{276}}, \bibinfo{pages}{1855} (\bibinfo{year}{1997}).

\bibitem[{\citenamefont{Kark}(2013)}]{kark2013ecotones}
\bibinfo{author}{\bibfnamefont{S.}~\bibnamefont{Kark}}, in
  \emph{\bibinfo{booktitle}{Ecological Systems}}
  (\bibinfo{publisher}{Springer}, \bibinfo{year}{2013}), pp.
  \bibinfo{pages}{147--160}.

\end{thebibliography}

\end{document}